\documentclass[aps,prd,superscriptaddress,11pt,showpacs,notitlepage]{revtex4}
	\usepackage{graphicx}
	\usepackage{amsmath,amssymb,mathrsfs}
\usepackage{epsf}
\usepackage{epsfig}

\usepackage{amsmath}

\usepackage{amssymb}

\numberwithin{equation}{section}

\newcommand{\be}{\begin{equation}}
\newcommand{\ee}{\end{equation}}
\newcommand{\bea}{\begin{eqnarray}}
\newcommand{\eea}{\end{eqnarray}}

\newcommand{\bb}{\bibitem}
 
\newcommand{\eqn}{\begin{eqnarray}}
\newcommand{\eqnx}{\end{eqnarray}}

\begin{document}
\title{Baryon chemical potential and in-medium properties of BPS skyrmions}
\author{C. Adam}
\affiliation{Departamento de F\'isica de Part\'iculas, Universidad de Santiago de Compostela and Instituto Galego de F\'isica de Altas Enerxias (IGFAE) E-15782 Santiago de Compostela, Spain}
\author{T. Kl\"{a}hn}
\affiliation{Institute for Theoretical Physics, University of Wroc\l aw, Poland}
\author{C. Naya}
\affiliation{Departamento de F\'isica de Part\'iculas, Universidad de Santiago de Compostela and Instituto Galego de F\'isica de Altas Enerxias (IGFAE) E-15782 Santiago de Compostela, Spain}
\author{J. Sanchez-Guillen}
\affiliation{Departamento de F\'isica de Part\'iculas, Universidad de Santiago de Compostela and Instituto Galego de F\'isica de Altas Enerxias (IGFAE) E-15782 Santiago de Compostela, Spain}
\author{R. Vazquez}
\affiliation{Departamento de F\'isica de Part\'iculas, Universidad de Santiago de Compostela and Instituto Galego de F\'isica de Altas Enerxias (IGFAE) E-15782 Santiago de Compostela, Spain}
\author{A. Wereszczynski}
\affiliation{Institute of Physics,  Jagiellonian University,
Lojasiewicza 11, Krak\'{o}w, Poland}

\begin{abstract}
We continue the investigation of thermodynamical properties of the BPS Skyrme model. 
In particular, we analytically compute the baryon chemical potential both in the full field theory and in a mean-field approximation. In the full field theory case, we find that the baryon chemical potential is always exactly proportional to the baryon density, for arbitrary solutions. 

We further find that, in the mean-field approximation, the BPS Skyrme model approaches the Walecka model in the limit of high density - their thermodynamical functions as well as the equation of state agree in this limit. This fact allows to read off some properties of the $\omega$-meson from the BPS Skyrme action, even though the latter model is entirely based on the (pionic) $SU(2)$ Skyrme field. On the other hand, at low densities, at the order of the usual nuclear matter density, the equations of state of the two models are no longer universal, such that a comparison depends on some model details. Still, also the BPS Skyrme model gives rise to nuclear saturation in this regime, leading, in fact, to an exact balance between repulsive and attractive forces. 

The perfect fluid aspects of the BPS Skyrme model, which, together with its BPS properties, form the base of our results, are shown to be in close formal analogy with the Eulerian formulation of relativistic fluid dynamics. Within this analogy, the BPS Skyrme model, in general, corresponds to a non-barotropic perfect fluid.

\end{abstract}
\maketitle 

%%%%%%%%%%%%%%%%%%%%%%%%%%%%%%%%%%%%%%%%%
\section{Introduction}
%%%%%%%%%%%%%%%%%%%%%%%%%%%%%%%%%%%%%%%%%
The derivation of properties of baryonic matter, especially beyond the nuclear matter density, is still one of the most challenging problems in current strong interaction physics. Since neither perturbative nor lattice computations apply in this regime, one is forced to use an effective model approach where both the field content and the form of the action are postulated from general considerations (symmetries, low energy degrees of freedom) rather than derived from the underlying fundamental quantum theory, i.e., QCD.

One of the most popular and successful effective theories is the Skyrme model framework \cite{skyrme}, where the fields are reduced to low energy effective chiral fields (pions and, optionally, higher mesons). Baryons, on the other hand, are not introduced as independent degrees of freedom but appear, instead,  as collective excitations in this mesonic matter, that is, as topological solitons with an identification between the baryon charge and the topological degree. 

Very promising results of the original Skyrme proposal when applied to the baryon sector \cite{sk}-\cite{scatter} (and also to some light nuclei) have to be contrasted with its problems in the modeling of higher nuclei and (infinite) nuclear matter. There are two main reasons for this fact. Firstly, the nuclear binding energies found in the Skyrme model are too large. Secondly, skyrmions with high baryon charge behave as crystals, that is, form a lattice built out of smaller charge substructures \cite{crystal}, which is in contrast to the liquid-type behavior of nuclear matter. Recently, three possible ways to cure the binding energy problem have been proposed: one may include a dominating sextic term in the model ({\it the near-BPS Skyrme model} \cite{BPS}, \cite{Marl}, \cite{Sp2}), add (infinitely) many vector mesons ({\it the BPS Skyrme vector meson model} \cite{SutBPS}, \cite{rho1}), or include a "repulsive" potential ({\it the lightly bound model} \cite{BPSM}). 

In particular, the {\it BPS Skyrme model} (the BPS restriction of the near-BPS Skyrme model) already provides quite accurate binding energies of the most abundant higher nuclei (after taking into account the semiclassical rotational and iso-rotational corrections as well as the Coulomb interaction and a small isospin breaking \cite{nearBPS}). It also leads to a perfect fluid energy-momentum tensor with {\it SDiff} symmetries. Therefore, this model plays the role of a field theoretical realisation of the liquid droplet model of atomic nuclei. Its near-BPS generalization, thus, seems to be a natural candidate for an effective model of nuclear matter within the Skyrme framework. Furthermore, the fact that the BPS Skyrme model has the energy-momentum tensor of a perfect fluid (without any averaging or mean-field approximation) led recently to a noticeable progress in the understanding of thermodynamic properties of skyrmions and nuclear matter as a skyrmionic medium \cite{term}. With our model, it was possible to find an exact equation of state relating the local energy density and the pressure which, as the energy density is generically spatially dependent (local spatial fluctuations), allows to study skyrmionic matter beyond the mean-field approximation. This was crucial for a better understanding of the long-standing issue of the too high compression modulus (too low compressibility) in the Skyrme model \cite{term}. The reason for these achievements is a rather unique property of the BPS Skyrme model, namely the equivalence of the microscopic (in terms of observables related to the original effective chiral fields) and macroscopic (in terms of thermodynamical functions and variables) thermodynamical descriptions. 

The sextic term, which is just the square of baryonic current, has been included in many effective models \cite{sextic} (see recent \cite{nitta}), and may be induced by the topological WZW coupling with the vector meson $\omega_\mu$ \cite{jackson}-\cite{ding}. It is known to significantly improve quantitative predictions of the model in the baryon sector.  

The fact that the BPS Skyrme model provides the rare possibility to study nuclear matter beyond the mean-field limit has been recently employed in a neutron star context, where the BPS Skyrme model was coupled to gravity \cite{star} (the original Skyrme model was used for the description of neutron stars in \cite{piette}). It has been found that such skyrmionic stars have masses and radii in very good agreement with current data for neutron stars, but, what is perhaps even more important, their properties do change if one performs the full field theory and gravity computation instead of the usual mean-field approximation known as the TOV approach. 

Beyond this progress, there is, however, another thermodynamical function (besides the energy (density), pressure and volume) which is extremely important but poorly understood in Skyrme type models. Namely, the baryon chemical potential, which is crucial if one wants to consider the model as a realistic description of cold and dense nuclear matter. It governs the stability of phases, possible phase transitions as well as the coexistence of different phases. Therefore, for any realistic application of gravitating BPS Skyrmions to neutron stars (which means taking into account $\beta$-equilibrium, existence of skin and crust or possible quark core - hybrid stars) one has to know this thermodynamical function. 

It is the main aim of the present work to fill this gap in our knowledge of the thermodynamical properties of the BPS Skyrme model. 

We shall find that it is again the geometrical nature of the BPS Skyrme model which allows to find the baryon chemical potential both in a mean-field approach and in an exact calculation, providing further evidence for the importance of geometric models of nuclear matter. Furthermore, using these results we are able to compute the mass of a BPS skyrmion in a skyrmionic medium, which is of high phenomenological importance as it may be related to the in-medium masses of baryons. 
%%%%%%%%%%%%%%%%%%%%%%%%%%%%%%%%%%%%%%%%%
\section{Properties of the BPS Skyrme model}
%%%%%%%%%%%%%%%%%%%%%%%%%%%%%%%%%%%%%%%%%
The BPS Skyrme model is defined by the following Lagrange density (we use the metric convention $(+,-,-,-)$)
\be
\mathcal{L}_{BPS} \equiv \mathcal{L}_6+\mathcal{L}_0 \equiv -(24\pi^2)^2 \lambda_6 \mathcal{B}_\mu \mathcal{B}^\mu- \lambda_0 \mathcal{U}.
\ee
It consists of the sextic term $\mathcal{L}_6$, i.e., the baryon current squared
\be
\mathcal{B}^\mu = \frac{1}{24\pi^2} \epsilon^{\mu \nu \rho \sigma} \mbox{Tr} \; L_\nu L_\rho L_\sigma, \;\;\; L_\mu = U^\dagger \partial_\mu U
\ee
and a non-derivative part, that is, a potential $\mathcal{U}(U)$. In the full near BPS Skyrme model, this BPS action is supplemented by the usual Skyrme (perturbative) model. However, in order to keep the binding energies on an acceptable level, the BPS part must give the leading contributions to masses, while the non-BPS part should enter as a rather small addition. Hence, for the description of static properties already the BPS Skyrme model should give a rather accurate approximation. For convenience, we redefine the coupling constants as $\lambda_6=\lambda^2 /(24)^2$ and $\lambda_0=\nu^2$.

Starting from now, we restrict considerations to the static case. The BPS property of the model means that one can reduce the static field equations to a first order (Bogomolny) equation
\be
\lambda \pi^2 \mathcal{B}_0= \pm \nu \sqrt{\mathcal{U}}
\ee
whose solutions saturate the following topological bound 
\be
E_{BPS} \geq 2\pi^2 \lambda \nu  |B| \langle\sqrt{\mathcal{U}} \rangle, \;\;\; \langle \sqrt{\mathcal{U}} \rangle \equiv \frac{1}{2\pi^2} \int_{\mathbb{S}^3} d\Omega \sqrt{\mathcal{U}} 
\ee
where $ \langle \sqrt{\mathcal{U}} \rangle$ is the average value of the square root of the potential on the target space. Further, $B=\int d^3 x {\cal B}^0$ is the baryon number (topological degree) of the Skyrme field. In fact, this equation can be analytically solved and possesses infinitely many $SDiff$ related solutions in each topological sector.

As already mentioned, the energy-momentum tensor has a perfect fluid form and, for static configurations, reads
\be
T^{00}=\varepsilon, \;\;\;  T^{ij}= - P \delta^{ij},  
\ee
where the energy density $\varepsilon$ and pressure $P$ are
\be
\varepsilon = \lambda^2 \pi^4 \mathcal{B}_0^2 +\nu^2 \mathcal{U},  \;\;\; P = \lambda^2 \pi^4 \mathcal{B}_0^2 -\nu^2 \mathcal{U} .
\ee
Hence, the Bogomolny equation is equivalent to the zero-pressure condition, \cite{bazeia}. Furthermore, the equation with a non-zero pressure gives rise to (non-equilibrium, one parameter) solutions of the full equations of motion, where from the conservation law for the energy-momentum tensor it follows that $P$ must be constant. Furthermore, the energy density - pressure equation of state (EoS) is not an algebraic equation but depends on the field component {\it (off-shell)}
\be
\varepsilon =P+2\nu^2 \mathcal{U}
\ee
or, after inserting an exact solution, on the space coordinates {\it (on-shell)}
\be
\varepsilon =\varepsilon (P, \vec{x})
\ee
where the particular dependence of the energy density on $\vec x$ depends both on the form of the potential and on the particular solution. This is a very peculiar property of the BPS Skyrme model: one can find an EoS (for the skyrmionic matter in the thermodynamical equilibrium) without any mean-field approximation i.e., generically, with a non-flat energy density distribution. In this sense, the BPS Skyrme model is a unique field theoretical tool which goes far beyond usual EFT. 

The fact that, generically, one deals with spatially non-constant energy densities poses, at first glance, a problem if one wants to compare the results of the BPS model with typical EoS obtained for other EFT (in a mean-field approximation). However, it is possible to derive a mean-field equation of state in the BPS Skyrme model, as well.  

Obviously, one may define an algebraic EoS which connects the total energy $E$ and pressure: $E=E(P)$ or geometrical volume $V$ and pressure: $V=V(P)$. Interestingly, by means of the Bogomolny equation, one can rewrite the total static energy and the geometric soliton volume as target space integrals independently of any particular solution and finds
\begin{equation}
E(P)=2\pi \lambda \nu |B|\tilde{E}, \;\;\; V(P)=2\pi |B| \frac{\lambda}{\nu} \tilde{V}
\end{equation}
where
\begin{equation} \label{E-prof}
\tilde{E}=\int_0^\pi d\xi \sin^2 \xi \frac{2\mathcal{U}+\tilde{P}}{\sqrt{\mathcal{U}+\tilde{P}}}, \;\;\; \tilde{V}(P)=\int_0^\pi d\xi \sin^2 \xi \frac{1}{\sqrt{\mathcal{U}+\tilde{P}}}
\end{equation}
or
\be
\tilde{E}=\frac{\pi}{2} \left\langle \frac{2\mathcal{U}+\tilde{P}}{\sqrt{\mathcal{U}+\tilde{P}}} \right\rangle, \;\;\;
\tilde{V}=\frac{\pi}{2} \left\langle \frac{1}{\sqrt{\mathcal{U}+\tilde{P}}} \right\rangle
\ee
and $\tilde{P}=P/\nu^2$. Here, in (\ref{E-prof})  we used the usual Skyrme field parametrization
\be \label{Sk-fields}
U=\cos \xi + i \sin \xi \; \vec{n} \cdot \vec{\tau} \; , \quad \vec n \equiv (\sin \Theta \cos \Phi , \sin \Theta \sin \Phi , \cos \Theta )
\ee
where $\vec{\tau}$ are Pauli matrices and $\vec n$ is a unit vector. Further we assumed that the potential depends on the Skyrme field only via the profile function $\xi$. 
It is sometimes useful to consider the axially symmetric ansatz for a skyrmion with baryon number $B$, where
\be \label{axi-sym}
\xi = \xi (r) \; , \quad \Theta = \theta \; , \quad \Phi = B \varphi
\ee
and $(r,\theta, \varphi)$ are spherical polar coordinates. Then the Bogomolny equation takes the form
\be
\frac{|B|\lambda}{2r^2} \sin^2 \xi \xi_r=- \nu \sqrt{\mathcal{U}+\tilde{P}}.
\ee
We want to emphasize, however, that all thermodynamic functions and variables as well as the thermodynamic relations between them are completely independent of any ansatz. They follow directly from the BPS equations (or, more generally, once integrated static field equations) and are the same for all solutions. They represent, therefore, properties of the model itself, and not of particular solutions.

The bulk observables $E(P)$ and $V(P)$ are given by the average values (on target space) of certain functions of the potential "shifted" by the pressure. They obey the thermodynamic relation
\be
P=-\frac{d E}{d V}
\ee
and, therefore, coincide with the corresponding thermodynamical functions, 
which means that for the BPS Skyrme model the microscopic (field theoretical) description is equivalent to the macroscopic approach (by thermodynamical functions). Despite the fact that, generically, we do not have an algebraic density-pressure EoS (which is an important fact as it allows to go beyond the mean-field approximation), one can perform such a limit and define an average energy density
\be
\bar{\varepsilon}=\frac{E}{V}
\ee
which obviously possesses an algebraic relation to the pressure (or volume). The fact that in the model we can compare mean-field with non-mean-field computations has been used recently for neutron stars \cite{star2}. In particular, we studied how their properties are modified by going beyond mean-field and taking into account the spatial dependence of the energy density of nuclear matter. In general, these two formulations (MF and exact) allow to investigate which predictions stem from the model itself, and which are related just to the mean field approximation.

Another important quantity is the particle number density which here is just the baryon charge density
\be
\rho_B=\mathcal{B}_0 .
\ee
Again, usually it has a non-constant (spatially dependent) form. An obvious proposition for an average particle (baryon) density is
\be
\bar{\rho}_B= \frac{B}{V}, \;\;\; B \equiv \int d^3 x \mathcal{B}_0 ,
\ee
as the total number of particles (the baryon charge) in a given volume $V$ is $B$. 

Although the energy density and pressure (or the particle density and pressure) are not related by an algebraic equations of state, there is an algebraic relation which connects all three of these local quantities, namely
\be
\varepsilon +P=2\lambda^2\pi^4 \rho_B^2 . \label{loc}
\ee
This equation will play a prominent role in the present work.
%%%%%%%%%%%%%%%%%%%%%%%%%%%%%%%%%%%%%%%%%
\section{Baryon chemical potential}
%%%%%%%%%%%%%%%%%%%%%%%%%%%%%%%%%%%%%%%%%
%%%%%%%%%%%%%%%%%%%%%%%%%%%%%%%%%%%%%%%%%
\subsection{Definition and properties}
%%%%%%%%%%%%%%%%%%%%%%%%%%%%%%%%%%%%%%%%%
The standard thermodynamical definition of the chemical potential is provided by the following relation with other thermodynamical functions
\be
\varepsilon +P=\mu \rho_B
\ee
From (\ref{loc}) it follows that
\be
\mu=2\lambda^2\pi^4 \rho_B.
\ee
Hence, the baryon chemical potential is proportional to the baryon charge density and, similarly to this quantity, it is a spatially non-trivial function. Let us remark that this is an off-shell, i.e., solution independent result. 

\vspace*{0.2cm}

Again, as for the energy density and particle number density, we may define a mean-field chemical potential. In general, it reads $\bar \mu =(\partial F/\partial N)$ where $F$ is the free energy, and $N$ is the particle number. In our zero temperature case we have $F=E$ and, further, the particle number $N$ is the baryon number $B$. We, therefore, get 
\begin{equation}
\bar{\mu} = \left( \frac{\partial E}{\partial B} \right)_V. \label{def}
\end{equation}
Of course, $\bar \mu$ must necessarily obey the same thermodynamical relation (for the averaged quantities)
\be
\bar \varepsilon + P= \bar{\mu} \bar{\rho}_B  . \label{glob}
\ee

This definition should be understood as follows. We consider a skyrmion in equilibrium, i.e., a solitonic solution with a given baryon charge $B_0$ which occupies a volume $V_0$ and has an energy $E_0$. Such a solution is a solution of the BPS equation and, therefore, corresponds to zero pressure $P=0$. Now, we want to check how the energy of a skyrmion varies if we increase its topological charge to $B=B_0+n$ but keep the volume fixed. Of course, this cannot be done in a smooth way, since the baryon charge is a topological, conserved quantity. Nonetheless, we may find a solution of the pressure equation which has increased topological charge $B$ and occupies the same volume $V_0$. The price we pay for that is the appearance of a non-zero pressure, which depends on the 'additional' baryon charge, $P=P(n)$.
Then we get two equations (here $V_0$ does not change, i.e., $V_0 (n,P) = V_0 (n=0,P=0)$)
\begin{equation}
E(n)=2\pi \lambda \nu (B_0+n)  \int_0^\pi d\xi \sin^2 \xi \frac{2\mathcal{U}+\tilde{P}}{\sqrt{\mathcal{U}+\tilde{P}}} \label{e}
\end{equation}
\begin{equation}
V_0=2\pi (B_0+n) \frac{\lambda}{\nu} \int_0^\pi d\xi \sin^2 \xi \frac{1}{\sqrt{\mathcal{U}+\tilde{P}}} . \label{V0}
\end{equation}
Using the definition (\ref{def}) of $\bar \mu$ and the fact that the volume remains constant we find
\be
\bar{\mu} = 2\pi \lambda \nu \frac{\pi}{2} \left[ \left< \frac{2\mathcal{U} +\tilde{P}}{\sqrt{\mathcal{U} +\tilde{P}}} \right> +\tilde{P} \left< \frac{1}{\sqrt{\mathcal{U} +\tilde{P}}} \right> \right] . \label{mu}
\ee 
This can be simplified to the following formula which expresses the chemical potential as a function of the pressure 
\be
\bar{\mu} = 4\pi \lambda \nu  \frac{\pi}{2} \left<  \sqrt{\mathcal{U} +\frac{P}{\nu^2}} \right> \label{mu sol} .
\ee 
On the other hand, equation (\ref{mu}) leads to the proper thermodynamical relation
\be
P= \bar{\mu} \bar{\rho}_B-\bar{\varepsilon}
\ee
which proves the consistency of our definition of the baryon chemical potential with standard thermodynamics. It is instructive to compare this formula with (\ref{loc}). If we integrate (\ref{loc}) over the soliton domain we get
\be
P V + E = 2\pi^4\lambda^2 \int d^3x \rho_B^2 .
\ee
Hence, the MF baryon chemical potential can be found in the following, alternative form
\be
\bar{\mu} = \frac{1}{B } 2\pi^4\lambda^2 \int d^3x \rho_B^2 \equiv 2\pi^4 \lambda^2 \frac{\int d^3x \rho_B^2 }{\int d^3x \rho_B }
\ee
which, after applying the Bogomolny equation, agrees with the target space average derived above. Obviously, $\bar{\mu} \neq 
2\lambda^2\pi^4 \bar{\rho}_B$, except for a very special case, that is, the step-function potential, for which all local quantities coincide with the corresponding average (MF) ones.

Furthermore, as
\be
\bar{\rho}_B=\frac{B}{V_0}= \frac{1}{2\pi} \frac{\nu}{\lambda} \frac{2}{\pi} \left< \frac{1}{\sqrt{\mathcal{U} +P/\nu^2 }} \right>^{-1}
\ee
we get
\be
\bar{\varepsilon} = \bar{\mu} \; \frac{1}{2\pi} \frac{\nu}{\lambda} \frac{2}{\pi} \left< \frac{1}{\sqrt{\mathcal{U} +P/\nu^2 }} \right>^{-1} -P \label{n sol}
\ee
which allows to express the energy density as a function of the chemical potential. 
Comparing now formulae (\ref{mu sol}) and (\ref{n sol}) we re-derive the well-known relation 
\be
\left( \frac{\partial \bar{\mu} }{\partial P} \right)_V = \frac{1}{\bar{\rho}_B}.
\ee
%%%%%%%%%%%%%%%%%%%%%%%%%%%%%%%%%%%%%%%%%
\subsection{High pressure limit}
%%%%%%%%%%%%%%%%%%%%%%%%%%%%%%%%%%%%%%%%%
In general, the baryon chemical potential - both the exact one  and its mean-field counterpart - is a complicated function of the particle number (baryon density), whose detailed form is governed by the used potential. The same is true for other thermodynamical quantities as well as the equation of state, which, too, depend on the potential. However, one can observe that at high pressure (equivalently energy or particle density) the model reveals a universal behaviour. Namely, it tends to the BPS Skyrme theory with the step-function potential. Indeed, for $P \gg \nu^2$ we get at leading order
\be
E=\pi^2 \lambda B \sqrt{P}
\ee
\be
V=\pi^2 B \frac{\lambda}{\sqrt{P}}.
\ee
Hence,
\be
\bar{\varepsilon}=P+B_\infty, \;\;\; \bar{\rho}_B = \frac{\sqrt{P}}{\pi^2\lambda}
\ee
and
\be
\bar{\mu}=2\pi^4 \lambda^2 \bar{\rho}_B
\ee
where $B_\infty$ is a bag constant at infinite pressure, see \cite{star2}. 
The fact that, asymptotically, the baryon chemical potential grows linearly with the baryon charge density (particle density) is generic for the BPS model and is not affected by a particular form of the potential. 

%%%%%%%%%%%%%%%%%%%%%%%%%%%%%%%%%%%%%%%%%
\subsection{Low pressure limit}
%%%%%%%%%%%%%%%%%%%%%%%%%%%%%%%%%%%%%%%%
At vanishing pressure the MF chemical potential is always equal to the equilibrium energy $E_0$ divided by the topological charge
\be
\bar{\mu}_0=\frac{E_0}{B}
\ee
while the exact chemical potential is, up to a multiplicative constant, the baryon density at equilibrium. 
This happens even in the non-compacton case i.e., when solitons are infinitely extended and the geometrical volume is infinite, which leads to zero average energy and particle density.  

For compact skyrmions, the model realizes a liquid-gas phase
transition. Indeed, at zero pressure and in the given (equilibrium)
volume $V_0$, we may have solutions with a smaller amount of 
topological charge by just removing some compactons from the given volume $V_0$. All these solutions are stable and form a collection of
compact solitons, surrounded by empty space. As they are BPS solutions ($P=0$), their total
energy is exactly proportional to the topological charge and, therefore,
$\bar{\mu} =\bar{\mu}_0$. Obviously, the corresponding MF energy
density and MF charge density will tend to 0 as the topological charge
decreases.
%%%%%%%%%%%%%%%%%%%%%%%%%%%%%%%%%%%%%%%%%
\subsection{Thermodynamic relations for specific potentials}
%%%%%%%%%%%%%%%%%%%%%%%%%%%%%%%%%%%%%%%%%
%%%%%%%%%%%%%%%%%%%%%%%%%%%%%%%%%%%%%%%%%
\subsubsection{The step function potential: $ \mathcal{U} = \Theta \; ( \mbox{Tr} \;(1-U ))$}
%%%%%%%%%%%%%%%%%%%%%%%%%%%%%%%%%%%%%%%%%
We start with a very special case - the step-function potential. This is a unique choice for the potential in the BPS Skyrme model which results in a constant energy density and particle number density (baryon charge density). Therefore, all local quantities completely agree with their mean-field (averaged) counterparts. 
The step-function potential reads 
\be
 \mathcal{U} = \Theta \; ( \mbox{Tr} \;(1-U)).
 \ee
The energy and pressure are constant 
\begin{equation}
\tilde{E} (\tilde{P}) = \frac{\pi}{2} \frac{2+\tilde{P}}{\sqrt{1+\tilde{P}}}, \;\;\; \tilde{V} (\tilde{P}) = \frac{\pi}{2} \frac{1}{\sqrt{1+\tilde{P}}}
\end{equation}
and
\begin{equation}
E(P)=2\pi \lambda \nu |B| \tilde{E}(\tilde{P}), \;\;\; V(P)=2\pi \frac{\lambda}{\nu} |B| \tilde{V} (\tilde{P}).
\end{equation}
Let us compute the baryon chemical potential from its definition (\ref{def}).
As before, we assume that we start in the equilibrium (where $P=0$) and then we add more particles (increase the baryon charge) keeping the volume constant. Let
\begin{equation}
E_0 = E(P=0)=2\pi^2 \lambda \nu |B_0|, \;\;\ V_0=V(P=0)=\pi^2 \frac{\lambda}{\nu} |B_0| .
\end{equation}
Then, after changing from $B_0$ to $B=B_0+n$ we get
\begin{equation}
V_0=\pi^2 \frac{\lambda}{\nu} B\frac{1}{\sqrt{1+\frac{P(B)}{\nu^2}}}
\end{equation}
\begin{equation}
E=\pi^2 \lambda \nu B \frac{2+\frac{P(B)}{\nu^2}}{\sqrt{1+\frac{P(B)}{\nu^2}}}
\end{equation}
where $P$ is due to the higher, than at the equilibrium $(B=B_0)$, topological number.  Both $E$ and $P$ are functions of $B$. Simplifying this expression we find
\be
E(P)=\nu^2 V_0 \left(2+\frac{P}{\nu^2} \right) \;\;\; \Rightarrow \;\;\; \left(\frac{\partial E }{\partial B}\right)_V = V_0 \left( \frac{\partial P}{\partial B} \right)_V .
\ee
However, from the constant volume condition we get
\be
\left( \frac{\partial P}{\partial B} \right)_V= \frac{2\pi^4 \lambda^2}{V_0^2} B
\ee
and
\be
\mu = 2\pi^4 \lambda^2 \frac{B}{V_0}= 2\pi^4\lambda^2 \bar{\rho}_B.
\ee
Similarly one can find
\be
P=\pi^4\lambda^2 \bar{\rho}_B^2-\nu^2 = \frac{1}{4\pi^4\lambda^2} \mu^2 - \nu^2
\ee
\be
\bar{\varepsilon}=\pi^4\lambda^2 \bar{\rho}_B^2+\nu^2 = \frac{1}{4\pi^4\lambda^2} \mu^2 + \nu^2
\ee
(see Fig. 1).
Obviously, the relation
\be
P+\bar{\varepsilon}=\mu \bar{\rho}_B
\ee
holds. 
%%%%%%%%%%%%%
\begin{figure}
\includegraphics[height=6cm]{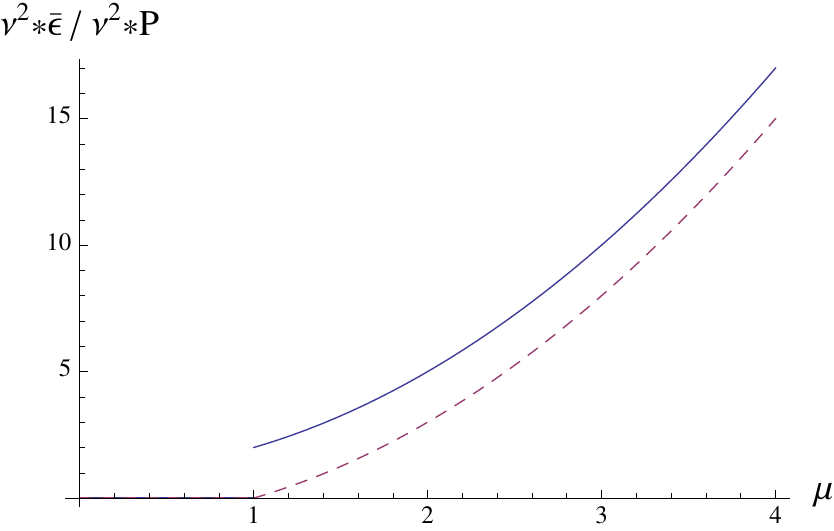}
\caption{Average energy density (continuous line) and pressure (dashed line) as functions of the chemical potential for the step function potential.}
\end{figure}
%%%%%%%%%%%%
Moreover we get the usual EoS
\begin{equation}
\bar{\varepsilon} = P + 2\nu^2 .
\end{equation}
As we see, the baryon chemical potential is always proportional to the particle density, not only asymptotically as has been proven in generality. Furthermore, the local and global chemical potential equations are exactly the same and $\mu=\bar{\mu}$. 
%%%%%%%%%%%%%%%%%%%%%%%%%%%%%%%%%%%%%%%%%
\subsubsection{No potential: $ \mathcal{U}=0$}
%%%%%%%%%%%%%%%%%%%%%%%%%%%%%%%%%%%%%%%%%
It is also possible to find the chemical potential for the BPS Skyrme model without any potential. Of course, at the equilibrium (no pressure) there are no stable soliton solutions, but for any non-zero $P$ skyrmions do exist. 
Then, 
\be
E(B) = \pi^2 \lambda B  \sqrt{P}, \;\;\; V=\pi^2 \lambda B \frac{1}{\sqrt{P}}
\ee
where we assume that we start with a given non-zero pressure solution with a fixed topological charge $B_0$ and then increase $B$ keeping the volume constant.  Hence,
\be
E(B)= \pi^4 \lambda^2 \frac{B^2}{V} \;\;\; \Rightarrow \;\;\ \mu = 2\pi^4\lambda^2 \frac{B}{V}
\ee
and
\be
\mu = 2\pi^4\lambda^2 \bar{\rho}_B
\ee
Moreover, as 
\be
P={\bar\varepsilon}= \pi^4\lambda^2 \bar{\rho}_B^2
\ee
we get that
\be
P=\bar{\varepsilon}= \frac{1}{4\pi^4\lambda^2} \mu^2
\ee
This case is quite similar to the step-function potential as the energy density and baryon density are again spatially constant. All quantities can be obtained from the former case by a simple $\nu \rightarrow 0$ limit. 
%%%%%%%%%%%%%%%%%%%%%%%%%%%%%%%%%%%%%%%%%
\subsubsection{Cubic potential: $\mathcal{U} =\frac{1}{2} (\xi -\frac{1}{2}\sin 2 \xi)$}
%%%%%%%%%%%%%%%%%%%%%%%%%%%%%%%%%%%%%%%%%
Let us now consider a more nontrivial situation, that is, a potential which leads to non-constant energy density. A simple example can be provided by the cubic (in the sense of the approach to the vacuum) potential $\mathcal{U} =\frac{1}{2} (\xi -\frac{1}{2}\sin 2 \xi$). This potential belongs to the so-called BPS potentials and provides exact and particularly simple solutions of the non-zero pressure integrals. Then,
\begin{equation}
\tilde{V}=2\left( \sqrt{\frac{\pi}{2} +\tilde{P}} -\sqrt{\tilde{P}}\right), \;\;\; \tilde{E}=\frac{1}{3} \left( 2\pi \sqrt{\frac{\pi}{2} +\tilde{P}}  -\tilde{P}\tilde{V}\right)
\end{equation}
or explicitly 
\begin{equation}
\tilde{E}=\frac{2}{3} \left( - \tilde{P} \left(\sqrt{\tilde{P}+\frac{\pi}{2}} -\sqrt{\tilde{P}}\right) +\pi \sqrt{\tilde{P}+\frac{\pi}{2}} \right) .
\end{equation}
Hence\begin{equation}
\bar{\varepsilon}=\frac{\nu^2}{3} \left( \pi \frac{\sqrt{\frac{\pi}{2} +\frac{P}{\nu^2}}}{\sqrt{\frac{\pi}{2} +\frac{P}{\nu^2}}- \sqrt{\frac{P}{\nu^2}}}  -\frac{P}{\nu^2}\right)
\end{equation}
Then, performing similar computations as before, we find how the energy and pressure vary if the baryon number is changed
\begin{equation}
P=\nu^2 \frac{\pi}{8} \frac{B^2}{B_0^2} \left( 1-\frac{B^2_0}{B^2} \right)^2 = \nu^2 \frac{\pi}{8} \frac{\bar{\rho}_B^2}{\bar{\rho}_{0,B}^{2}} \left( 1-\frac{\bar{\rho}_{0,B}^2}{\bar{\rho}^2_{B}} \right)^2 
\end{equation}
\begin{equation}
E= \frac{2\pi^{5/2}}{3\sqrt{2}} B_0 \lambda \nu \left( 1+\frac{B^2}{B_0^2} - \frac{1}{4} \frac{B^2_0}{B^2} \left( 1- \frac{B^2}{B_0^2} \right)^2 \right)
\end{equation}
Here $\bar{\rho}_{0,B}$ is the average baryon density at equilibrium. Therefore, 
\begin{equation}
\bar{\mu} = \frac{\pi^{5/2}}{3\sqrt{2}} \lambda \nu  \left( 3 \frac{B}{B_0}+ \frac{B^3_0}{B^3}  \right) = \frac{\pi^{5/2}}{3\sqrt{2}} \lambda \nu  \left( 3 \frac{\bar{\rho}_B}{\bar{\rho}_{0,B}} + \frac{\bar{\rho}_{0,B}^3}{\bar{\rho}_{B}^{3}} \right) 
\end{equation}
and 
\begin{equation}
\bar{\varepsilon} = \frac{\pi }{6}  \nu^2 \left( 1+  \frac{B^2}{B_0^2} - \frac{1}{4}  \frac{B_0^2}{B^2} \left(1+  \frac{B^2}{B_0^2} \right)^2\right) =
 \frac{\pi }{6}  \nu^2 \left( 1+  \frac{\bar{\rho}_B^2}{\bar{\rho}_{0,B}^2}- \frac{1}{4}  \frac{\bar{\rho}_{0,B}^2}{\bar{\rho}^2_{B}}\left(1+  \frac{\bar{\rho}_{B}^2}{\bar{\rho}^2_{0,B}} \right)^2\right) 
\end{equation}
(see Figs. 2,3).
%%%%%%%%%%%%%%%%%
\begin{figure}
\includegraphics[height=5cm]{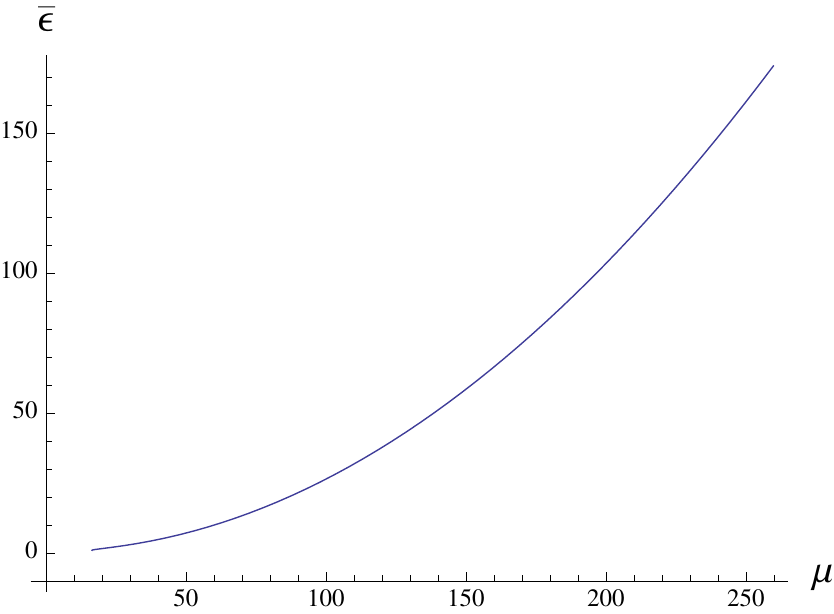}
\includegraphics[height=5cm]{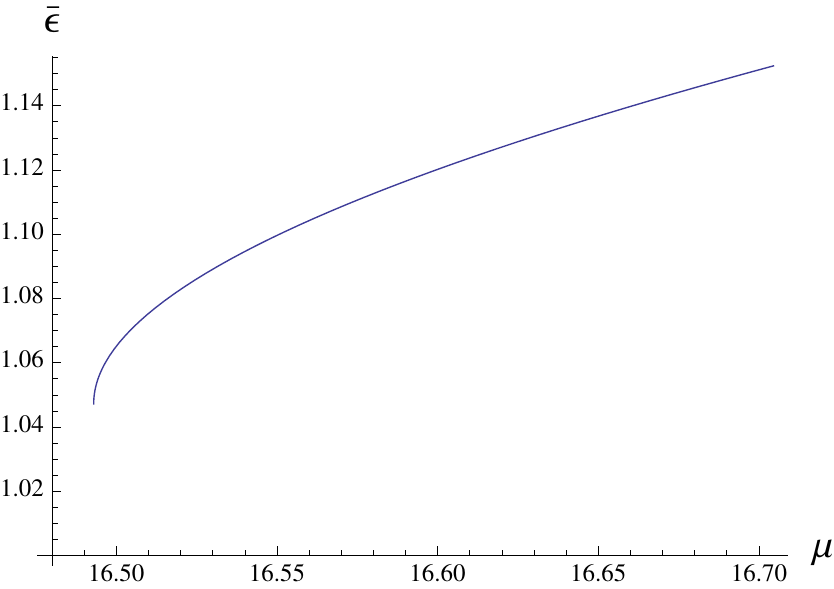}
\caption{Left figure: Average density as a function of the MF chemical potential for the cubic potential $\mathcal{U}=\frac{1}{2} (\xi -\frac{1}{2}\sin 2 \xi)$. Here $\nu^2=\lambda=1$. Right figure: zoom-in close to saturation density. }
\end{figure}
%%%%%%%%%%%%%%%%
\begin{figure}
\includegraphics[height=6cm]{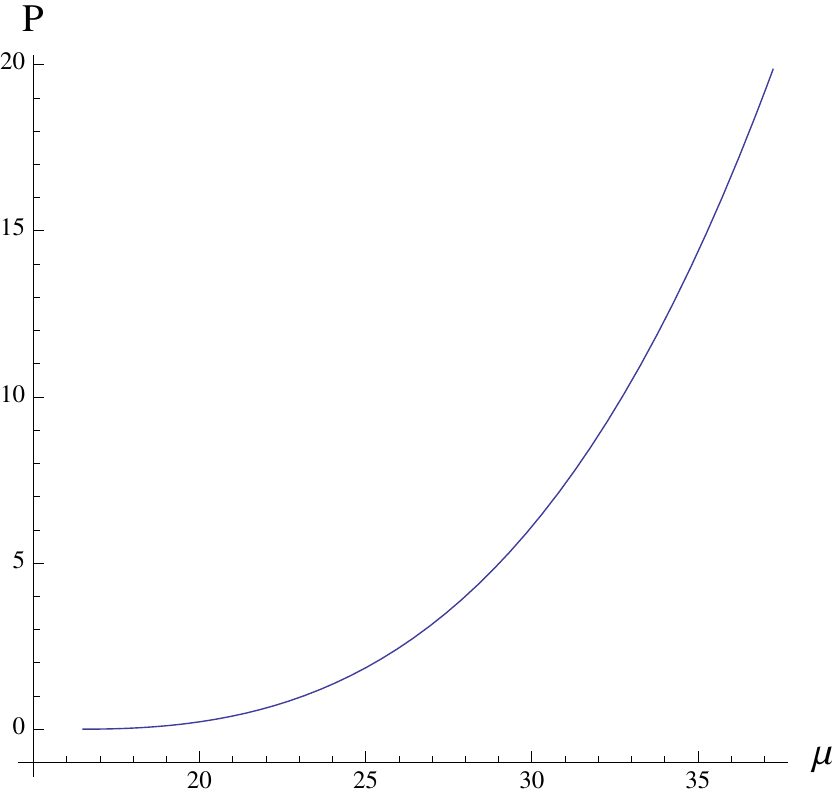}
\caption{Pressure as a function of the MF chemical potential for  the cubic potential $\mathcal{U}=\frac{1}{2} (\xi -\frac{1}{2}\sin 2 \xi)$. Here $\nu^2=\lambda=1$}
\end{figure}
%%%%%%%%%%%%%%%%%%
Asymptotically, for sufficiently high chemical potential we get
\begin{equation}
\bar{\mu} = \frac{\pi^{5/2}}{\sqrt{2}} \lambda \nu \frac{B}{B_0}
\end{equation}
and
\begin{equation}
P=\frac{\pi}{8} \left( \frac{B}{B_0} \right)^2 \nu^2, \;\;\; \bar{\varepsilon}=\frac{\pi}{8} \left( \frac{B}{B_0} \right)^2 \nu^2
\end{equation}
or
\begin{equation}
P=\frac{1}{4\pi^4 \lambda^2}  \bar{\mu}^2, \;\;\; \bar{\varepsilon}=\frac{1}{4\pi^4 \lambda^2}  \bar{\mu}^2,
\end{equation}
Hence, for high value of the chemical potential we as always obtain the linear EoS
\begin{equation}
P=\bar{\varepsilon}
\end{equation}
where the subleading constant $B_\infty$ has been omitted. 
%%%%%%%%%%%%%%%%%%%%%%%%%%%%%%%%%%%%%%%%%
\subsubsection{Non-compacton potential: $\mathcal{U} =\frac{1}{4} (\xi -\frac{1}{2}\sin 2 \xi)^2$}
%%%%%%%%%%%%%%%%%%%%%%%%%%%%%%%%%%%%%%%%%
Another potential we want to discuss is a potential with a sextic approach to the vacuum. It means that BPS skyrmions are no longer compactons but usual, infinitely extended solitons. It results in an infinite geometrical volume at the equilibrium. Then, the average energy density goes to zero at the equilibrium. Hence, similar as in the non-potential case, we are forced to close the skyrmionic medium in a given, finite volume by the application of a nonzero external pressure. Then, we find
\be
V=2\pi \frac{\lambda}{\nu} B \; \mbox{arsinh} \frac{\pi \nu}{2\sqrt{P}}, \;\;\; E=\pi^2 \lambda \nu B  \sqrt{\frac{P^2}{\nu} +\frac{\pi^2}{4}}
\ee
Thus, in terms of the baryon number
\be
P=\frac{\pi^2\nu^2}{4} \frac{1}{\sinh^2 \frac{\nu}{2\pi \lambda} \frac{V}{B}
}, \;\;\; E= \frac{\pi^3}{2} \lambda \nu  B \coth \left(  \frac{\nu}{2\pi \lambda}
\frac{V}{B} \right)
\ee
or using the average particle density
\be
P=\frac{\pi^2\nu^2}{4} \frac{1}{\sinh^2 \frac{\nu}{2\pi \lambda \bar{\rho}_B} 
}, \;\;\; E= \frac{\pi^3}{2} \lambda \nu  B \coth \left(  \frac{\nu}{2\pi \lambda \bar{\rho}_B}
\right) .
\ee
Therefore, the MF chemical potential reads
\be
\bar{\mu} = \frac{\pi^3}{2} \lambda \nu \left[ \coth \left(  \frac{\nu}{2\pi \lambda \bar{\rho}_B}
\right) + \frac{\nu}{2\pi \lambda \bar{\rho}_B} \frac{1}{\sinh^2 \frac{\nu}{2\pi \lambda \bar{\rho}_B} 
} \right]
\ee
For high particle density we find the usual linear relation
\be
\bar{\mu} = 2\pi^4 \lambda^2 \bar{\rho}_B, \;\;\;  \bar{\rho}_B \rightarrow \infty .
\ee
Here it is also possible to derive the MF chemical potential for vanishing particle density
\be
\bar{\mu} =  \frac{\pi^3}{2} \lambda \nu+ \pi^2 \nu^2 \frac{1}{\bar{\rho}_B} e^{-\frac{\nu}{\pi \lambda} \frac{1}{\bar{\rho}_B }}, \;\;\;  \bar{\rho}_B \rightarrow 0 .
\ee
Since the average mean-field energy density goes to zero at the equilibrium ($P=0$), this case, at least from the mean-field perspective, does not look like a bag type model. The saturation density is simply zero.  
%%%%%%%%%%%%%%%%%%%%%%%%%%%%%%%%%%%%%%%%%
\subsubsection{The Skyrme potential: $ \mathcal{U}_\pi = \frac{1}{2}\mbox{Tr} \; (1-U )$}
%%%%%%%%%%%%%%%%%%%%%%%%%%%%%%%%%%%%%%%%%
Finally, we want to present the result for the usual (pion mass) Skyrme potential
\be
\mathcal{U}=\mathcal{U}_\pi=\frac{1}{2} \mbox{Tr} \; (1-U)= 2 \sin^2 \frac{\xi}{2} .
\ee
Then, the MF baryon chemical potential as a function of the average baryon density is implicitly given by the following formulae
\be
\bar{\mu}= 4\pi \lambda \nu \frac{4}{15 \nu^4}  \sqrt{2+\frac{P}{\nu^2}} \left( (4 \nu^4 +2 \nu^2P +P^2) E \left[ \frac{2\nu^2}{2\nu^2 +P}\right] -
P (\nu^2 +P)K \left[\frac{2\nu^2}{2\nu^2 +P} \right]  \right)
\ee
\be
\bar{\rho}_B= \frac{3\nu}{8 \pi \lambda} \frac{ \nu^2 }{\sqrt{2+\frac{P}{\nu^2}} \left( (\nu^2+P)E\left[ \frac{2\nu^2}{2\nu^2+P}\right] -P K\left[\frac{2\nu^2}{2\nu^2+P} \right] \right)}
\ee
Here, $E$ and $K$ are the complete elliptic integrals of the first and second kind, respectively. 
%%%%%%%%%%%%%%%%%%%%%%%%%%%%%%%%%%%%%%%%%
\section{In-medium skyrmions}
%%%%%%%%%%%%%%%%%%%%%%%%%%%%%%%%%%%%%%%%%
Using the framework presented in this paper, it is also possible to obtain energies (masses) of skyrmions in the skyrmionic medium. This is of high importance, as it allows to find in-medium masses of baryons (nucleons). 
In order to find how the energy of a skyrmion with a given topological charge $B_0$ varies if it is immersed in the skyrmionic medium, we consider the following integral
\begin{equation}
E_{B_0}(n)=2\pi \lambda \nu B_0  \int_0^\pi d\xi \sin^2 \xi \frac{2\mathcal{U}+\tilde{P}}{\sqrt{\mathcal{U}+\tilde{P}}} . \label{inmed}
\end{equation}
This is the energy of charge $B_0$ skyrmion under the external pressure $P$. This pressure is induced by the additional baryon charge $n$ injected into the equilibrium solution (volume $V_0$, pressure $P=0$) without changing the volume
\begin{equation}
V_0=2\pi (B_0+n) \frac{\lambda}{\nu} \int_0^\pi d\xi \sin^2 \xi \frac{1}{\sqrt{\mathcal{U}+\tilde{P}}} . \label{V0}
\end{equation}
If compared with equation (\ref{e}), the energy differs only by the change of the overall multiplicative factor $B_0+n$ into $B_0$. This corresponds to the fact that  now we are interested in the energy of the charge $B_0$ baryon surrounded by a skyrmionic medium with the additional charge $n$, which leads to a non-zero pressure. Of course, this pressure is the same in the medium and in the original (now compressed) skyrmion.
Therefore, the energy (\ref{inmed}) is, in fact, the in-medium energy of the charge $B_0$ skyrmion. Using these formulas and the expression for the baryon chemical potential, it is possible to express the energy of a skyrmion as a function of the mean-field chemical potential.  

As an example, we consider the step function potential. Then, one can show that
\be
E_{B_0}(n)=\pi^2 \lambda \nu (B_0+n)\left( 1+\frac{B_0^2}{(B_0+n)^2}\right)
\ee
which, after taking into account the formula for the baryon chemical potential, can be rewritten as
\be \label{in-med-E}
E_{B_0}(\mu)= \frac{B_0}{2} \mu \left( 1+\frac{4\pi^4\lambda^2\nu^2}{\mu^2} \right)
\ee
(remember $\mu \equiv \bar \mu$ for the step function potential).
This is valid for $\mu \ge \mu_0$ i.e., above equilibrium. For $\mu = \mu_0$, i.e., from the vacuum value until the equilibrium, the in-medium mass is always the same and equal to the equilibrium mass. To show this, let us assume that we start with a collection of charge one skyrmions with the total baryon charge $B_0$, which in equilibrium occupy the volume $V_0$. Now, we take away charge one skyrmions one after the other. Obviously, due to the contact form of the interaction and the BPS nature of the solutions, each removed unit skyrmion has the same energy (mass). However, the density of the medium, which is now a gas of BPS skyrmions in the fixed volume $V_0$,  decreases. 

Repeating the computation from the previous section, we can conclude that for any potential asymptotically the in-medium energy of a skyrmion (baryon) always behaves as
\be
E_{B_0}(\mu)= \frac{B_0}{2} \mu \;\;\;\; \mbox{at} \;\;\;\; \mu \rightarrow \infty.
\ee
In general (arbitrary potential), the in-medium energy of charge one baryon reads
\be
E_{B=1} =\bar{\mu} - \frac{PV_0}{1+n} = \bar{\mu}-\frac{P}{\bar{\rho}_B} ,
\ee
where the second part vanishes at the equilibrium $(P=0)$ and tends to $\frac{1}{2}\bar{\mu}$ at asymptotically large densities. Let us underline that this result is {\it beyond} the mean-field (constant energy density) approximation. 

In the full near-BPS Skyrme model or in the BPS Skyrme model with semiclassical contributions included, it is reasonable to expect a modification of the obtained in-medium mass dependence. Indeed, due to non zero binding energies one may expect a skyrmion mass which increases with decreasing medium density at small densities. At sufficiently high density, the universal relation $E_{B_0} \sim \frac{B_0}{2} \mu $ should again be valid.

In an analogous manner, we may compute the in-medium size of a BPS skyrmion. In general, the volume $V_0$ occupied by the original charge $B_0$ soliton at equilibrium is reduced by adding the additional topological charge $n$ i.e., by increasing the medium density 
\be \label{in-med-V}
V_{B_0}=\frac{B_0}{B_0+n} V_0= \frac{B_0}{\bar{\rho}_B} \rightarrow  \frac{2\pi^4\lambda^2}{\bar{\mu}} B_0 \;\;\; \mbox{at} \;\; \bar{\mu} \rightarrow \infty ,
\ee
and the radius of the compacton reads
\be
R_{B_0}=\left( \frac{3 B_0}{4\pi \bar{\rho}_B}\right)^{1/3} \rightarrow \; \left( \frac{3\pi^2\lambda^2}{2} \right)^{1/3} \bar{\mu}^{-1/3} \;\;\; \mbox{at} \;\; \bar{\mu} \rightarrow \infty .
\ee
Observe that, due to the thermodynamical properties of the BPS Skyrme model, the global quantities (the mean-field energy density, the mean-field particle density, the mean-field chemical potential) of a skyrmion and the surrounding medium always coincide. On the other hand, their local counterparts differ. 
%%%%%%%%%%%%%%%%%%%%%%%%%%%%%%%%%%%%%%%%%
\section{Physical implications}
%%%%%%%%%%%%%%%%%%%%%%%%%%%%%%%%%%%%%%%%%
So far, we have presented some thermodynamical properties of the BPS Skyrme model from a more theoretical point of view. In this section, we want to study their effect on physical properties of baryons and nuclei if described by BPS skyrmions. Furthermore, we shall use our results to relate the BPS Skyrme model to effective theories including $\omega$ and $\sigma$ mesons and show how their properties are hidden in our skyrmionic (pionic) action, with an exact balance between attractive and repulsive channels. 
%%%%%%%%%%%%%%%%%%%%%%%%%%%%%%%%%%%%%%%%%
\subsection{The Walecka model and BPS skyrmions}
%%%%%%%%%%%%%%%%%%%%%%%%%%%%%%%%%%%%%%%%%
The Walecka model of nuclear matter consists of nucleons (neutron and proton spinors) which interact with the scalar $\sigma$ meson and the vector $\omega$ meson \cite{walecka}
\be
\mathcal{L}_W= \mathcal{L}_N+\mathcal{L}_{\sigma, \omega} + \mathcal{L}_{int}
\ee
where
\be
\mathcal{L}_N=\bar{\psi} \left( i\gamma^\mu \partial_\mu -m_N + \mu \gamma^0\right) \psi
\ee
\be
\mathcal{L}_{\sigma, \omega} = \frac{1}{2} (\partial_\mu \sigma)^2 - \frac{1}{2} m_\sigma^2 \sigma^2  - \frac{1}{4} \omega_{\mu \nu} \omega^{\mu \nu} + \frac{1}{2} m_\omega^2 \omega_\mu \omega^\mu
\ee
\be
\mathcal{L}_{int}= g_\sigma \bar{\psi} \sigma \psi +g_\omega \bar{\psi} \gamma^\mu \omega_\mu \psi.
\ee
Here $\omega_{\mu \nu} = \partial_\mu \omega_\nu-\partial_\nu \omega_\mu$. A self-interaction term for the scalar meson may also be included. The mean-field approximation means that one computes the partition function (in the thermodynamical limit) 
\be
Z= \int \mathcal{D} \bar{\psi} \mathcal{D} \psi \mathcal{D} \sigma \mathcal{D} \omega \;  e^{\int \mathcal{L}_W}
\ee
in the limit where the bosonic fields are approximated by their constant condensate values $\bar{\sigma}$, $\bar{\omega}_0$. Then all derivative dependent terms disappear and  the interactions are simplified to a mesonic background field seen by the nucleons. This means that we arrive at a free fermion model with shifted parameters,
\be
\mathcal{L}_W=\bar{\psi} \left( i\gamma^\mu \partial_\mu -m_N^* + \mu^* \gamma^0\right) \psi - \frac{1}{2} m_\sigma^2 \bar{\sigma}^2 + \frac{1}{2} m_\omega^2 \bar{\omega}_0^2
\ee
where
\be \label{walecka-nuc-mass}
m^*_N=m_N-g_\sigma \bar{\sigma}, \;\;\; \mu^*=\mu - g_\omega \bar{\omega}_0 .
\ee
One should remember that the baryon chemical potential (which enters in all thermodynamical relations) is still $\mu$. However, the effective chemical potential $\mu^*$ sets the Fermi energy of the "effective" free fermions
\be
E_F^*=\mu^*=\sqrt{k_F^2+(m_N^*)^2} .
\ee
Here we assume the zero temperature limit. The baryon density reads \cite{nuclmat}
\be
\rho_B=\frac{2k_F^3}{3\pi^2}
\ee
and is related to the $\omega$ meson vacuum value by
\be
\bar{\omega}_0=\frac{g_\omega}{m_\omega^2} \rho_B .
\ee
An interesting observation is that in the limit of large density, i.e., $k_F \rightarrow \infty$, the equation of state in the Walecka model exactly coincides with the MF EoS derived for the BPS Skyrme model at pressures $P \gg B_\infty$ (which in practice means $P \gg \nu^2$). Namely 
\be
P=\varepsilon .
\ee
Moreover, in the Walecka model this limit reads as
\be
\varepsilon = \frac{1}{2} \frac{g^2_\omega}{m_\omega^2} \rho_B^2 .
\ee
Comparing this with the universal relation between average energy density and baryon density in the MF BPS Skyrme model we find that
\be
\pi^4\lambda^2 =  \frac{1}{2} \frac{g^2_\omega}{m_\omega^2} . \label{lambda}
\ee
Another important observation can be made if we analyse the large density limit for the effective chemical potential. Then we find
\be
 k_F \rightarrow \infty \;\;\; \Rightarrow \;\;\; \mu^*=k_F \;\; \mbox{and} \;\;\; \mu^* \sim \rho_B^{1/3}
 \ee
 Hence, at $k_F \rightarrow \infty$
 \be
 \mu = \mu^*+g_\omega \bar{\omega}_0 = \mu^* + \frac{g^2_\omega}{m_\omega^2} \rho_B \sim \frac{g^2_\omega}{m_\omega^2} \rho_B .
 \ee
 This formula has an exact counterpart in the BPS Skyrme model in the MF approach and in the high density limit
 \be
 \bar{\mu} = 2\pi^4 \lambda^2 \bar{\rho}_B .
 \ee
 Comparing the last two expression, we find independently again the relation (\ref{lambda}). 
 It is a striking fact that in the exact (non-mean field) microscopic thermodynamics in the BPS Skyrme model this formula is valid at any pressure (density). Indeed, as we know
\be
\mu = 2\pi^4 \lambda^2 \rho_B.
\ee
 
The conclusion is that the BPS Skyrme model and the Walecka model are equivalent in the high density regime, as far as thermodynamical properties are concerned and a mean-field approximation is made.  The obtained relation between the BPS Skyrme and Walecka model parameters allows us to read off the $\omega$ meson coupling constant in the BPS Skyrme model. For $m_\omega=738 $ MeV and $g_\omega^2/4\pi=10-12$, which is the empirical value of the $\omega$-nucleon coupling, we get the acceptable values for the Skyrme model parameter $\lambda^2 = 9-11$ MeV\;fm$^3$. It may be also compared with an upper bound for the coupling constant $g_\omega = 25.4$ \cite{AN}. Then, $\lambda^2 \le 47$ MeV\;fm$^3$. Let us remark that all previously used potentials led to $\lambda^2$ slightly bigger than the optimal value, but significantly below the upper bound \cite{nearBPS,star,star2}. Undoubtedly, this bound on the BPS Skyrme model parameter may help to constrain the potential part of the action. 
 
The reason why these two models possess the same thermodynamical properties at high density limit originates from the fact that in this regime the Walecka model is dominated by the vector meson sector. From the interaction point of view, the vector meson couples to the baryon current and effectively the dominating part of the Lagrangian is the baryon current squared. But this is exactly the derivative part of the BPS Skyrme model, although the latter is expressed in a topological manner and not by nucleon spinors. We comment that the same effect shows up in a quark bag model with a vector interaction \cite{thomas}. In any case, it is a rather interesting observation that it is possible to read-off the ratio between the $\omega$ meson mass and its coupling constant to the nucleon in the BPS Skyrme model, even though there are no obvious $\omega$ meson degrees of freedom in its action. Instead, we only have the usual pionic fields. The $\omega$ meson is {\it hidden} in the form of the BPS action. 
We remark that the sextic term of the BPS Skyrme model may also be obtained from a Skyrme type model with the $\omega$ meson included explicitly \cite{AN}, as the leading contribution in a derivative expansion \cite{jackson}. This leads to the plausible conjecture that soliton solutions in a Skyrme model with $\omega $ mesons should be quite similar to soliton solutions in a Skyrme model with the sextic term (the baryon current squared) included. That this is indeed the case has been demonstrated recently numerically in the baby Skyrme model in one lower dimension  \cite{sutcliffe2}, where the baryon current squared is identical to the Skyrme term.

At lower densities, the EoS of both models get much more complicated, and the BPS Skyrme model EoS now depends on the potential, such that a direct comparison is no longer obvious. In addition, at low densities the field theoretical picture of the BPS Skyrme model is probably more adequate  than the mean-field approximation. For example, we know that the constant energy density approximation is not the proper one if one wants to derive the compressibility of skyrmionic matter at equilibrium \cite{term}. 

\vspace*{0.1cm}
It is interesting to note that the {\it local} chemical potential is exactly linear in the baryon density for the whole range of the pressure, not just in the asymptotic regime, as happens in the Walecka model. 
%%It is also quite amazing that the equivalence between the BPS Skyrme model and the $\omega$ meson occurs for any range of pressure 
%%(density) if one considers the {\it local} chemical potential. Then, the proportionality to the particle (baryon) density is exact - not only in the %%%asymptotic regime. 
Besides, one may perhaps expect that there is a relation between the equation of state in the BPS model and a particular form of potential for the $\sigma$ meson field. 

%%%%%%%%%%%%%%%%%%%%%%%%%%%%%%%%%%%%%%%%%
\subsection{In-medium mass of nucleons}
%%%%%%%%%%%%%%%%%%%%%%%%%%%%%%%%%%%%%%%%%
Now we want to identify nucleons with charge one skyrmions, in order to see what results we get from our in-medium skyrmion mass calculations for the corresponding in-medium masses of nucleons. As we have shown, the mass of a charge one skyrmion is density independent as long as the medium density does not exceed the saturation density, i.e., the energy density at equilibrium. Above this density, it starts to grow with an asymptotically linear dependence $M\sim \frac{\bar{\mu}}{2}$. This behaviour agrees with the in-medium mass behaviour recently found using a holographic approach \cite{inmedium-hol}, where the mass of nucleons decreases from its vacuum value until one reaches the saturation point, where it begins to grow. 
If we want to compare with further approaches, the following subtlety must be taken into account. By construction, our in-medium nucleon masses are always the total in-medium skyrmion (rest) energies per baryon number. This is not the case in several approaches (e.g., in the Walecka model, see Eq. (\ref{walecka-nuc-mass}), or in \cite{rho-med}), where the in-medium mass of the nucleon is exclusively induced by in-medium changes of certain coupling constants, which may even lead to an in-medium reduction of the nucleon mass above nuclear saturation. In these approaches, further contributions to the total energy stem from in-medium modified nucleon-nucleon interactions, which may give significant contributions to the total in-medium energy per baryon number. Obviously, our results must always be compared with these total energies per baryon number.  
%%On the other hand, such an in-medium mass growth above the saturation point (or any other 'critical' density point) should be contrasted with the %%recent finding in the hidden local symmetry Lagrangian  \cite{rho-med} where a density independent mass regime is found beyond a point with %%%half-skyrmionic phase generation. 

Within our framework, the reduction of the nucleon mass between the vacuum and saturation density expected on general physical grounds, may probably be obtained by an extension of the model to its near-BPS generalization by taking into account the perturbative part of the Skyrme theory (pionic perturbative part) and/or by the semiclassical corrections. 

Let us also remark that the in-medium properties of skyrmions in the present work, in contrast to the in-medium modified Skyrme Lagrangian \cite{yak2}, have been obtained within the BPS Skyrme model without introducing any medium-dependent constants. Such a medium modified BPS Skyrme model would have the following form
\be
\tilde{\mathcal{L}}_{BPS} \equiv  - \pi^4 \bar\lambda^2 (\mu, \vec{x}) \mathcal{B}_\sigma \mathcal{B}^\sigma- \bar \nu^2(\mu, \vec{x}) \mathcal{U}.
\ee
Using our results one may try, however, to express the coupling constants $\bar\lambda$ and $\bar\nu$ as functions of the medium density $\mu$ (and, probably, of the coordinates $\vec x$) and fit to the correct in-medium mass dependence. Then one may check how good this approximation is by comparing with other thermodynamical properties. At this stage of research, we shall just discuss the step-function potential case, leaving other more interesting potentials for the future.  For this potential, all local densities are equal to their mean-field counterparts and, therefore, also the in-medium constants $\bar\lambda=\bar \lambda(\mu)$, $\bar \nu = \bar \nu(\mu)$ are expected to be spatially constant. 
Indeed, using the equilibrium (no medium) expressions $V_0 \equiv V_{B_0}(\mu_0) = \pi^2 B_0 (\lambda /\nu)$ and $E_0 \equiv E_{B_0} (\mu_0) = 2\pi^2 \lambda \nu B_0 \equiv \mu_0 B_0$ together with their in-medium values (see Eqs. (\ref{in-med-E}) and (\ref{in-med-V}))
\be
E_{B_0} (\mu) = E_{B_0} (\mu_0) \frac{1}{2} \left( \frac{\mu}{\mu_0} + \frac{\mu_0}{\mu} \right) \; , \quad V_{B_0} (\mu) = V_{B_0} (\mu_0) \frac{\mu_0}{\mu}
\ee
we find the in-medium coupling constants
\be
\bar \lambda ^2= \lambda^2 \, \frac{1}{2} \left( \frac{\mu_0^2}{\mu^2} + 1 \right) \; , \quad \bar \nu^2 = \nu^2 \,
\frac{1}{2} \left( \frac{\mu^2}{\mu_0^2} + 1 \right) .
\ee 
Here we consider the liquid phase, i.e., $\mu \geq \mu_0$. In the
gaseous phase, where the energy density as well as the particle
density approach 0, the chemical potential has always the same value
$\mu_0$. Similarly, the in-medium skyrmion mass and its volume remain
unchanged.

It should also be noticed that typical situations where such medium modified Skyrme Lagrangians are considered are nucleons in atomic nuclei or in infinite nuclear matter at {\it equilibrium}. As we commented before, in this regime the BPS Skyrme model does not differ from its in-vacuum version. To make predictions for this regime, we have to include semiclassical corrections or extend the model to the near-BPS one. One the other hand, usually the coupling functions in the in-medium generalized Skyrme model are assumed to be spatially constant for infinite nuclear matter \cite{yak2}. Here, the BPS Skyrme model can be of some help, at least for sufficiently flat potentials, because the original Skyrme model with its crystalline structure for large $B$ definitely does not lead to a flat energy density.

%%%%%%%%%%%%%%%%%%%%%%%%%%%%%%%%%%%%%%%%%
\subsection{The BPS Skyrme model, $\sigma$ and $\omega$ mesons, and chiral symmetry}
%%%%%%%%%%%%%%%%%%%%%%%%%%%%%%%%%%%%%%%%%
Usually, in the Walecka model (or other, more general low energy effective models) the equilibrium at nuclear saturation is the result of a precise balance between the repulsive forces induced by the $\omega$ meson and the attractive forces due to the $\sigma$ mesons (plus some small contributions of further mesons in more general models).
It is part of the elegance of the BPS Skyrme model that, as a result of the BPS equation, it provides an {\it exact cancelation} between these forces, without any need for a fine-tuning of coupling constants, which leads to an exact equilibrium at the nuclear saturation density, and to exactly zero classical binding energies. 
Indeed, the static energy functional $E=E_6 + E_0 =\int d^3 x (\varepsilon_6 + \varepsilon_0)$, with  $\varepsilon_6 = \pi^4 \lambda^2 {\cal B}_0^2 $ and $\varepsilon_0 =  \nu^2 {\cal U}$, consists of two terms which scale oppositely under Derrick scaling $\vec x \to \Lambda \vec x$. Concretely, $E_6$ tends to expand the field configuration, inducing a repulsive force between different volume elements of the soliton, whereas the potential part $E_0$ tends to collapse the configuration, corresponding to an attractive force. 
It is instructive to consider the resulting BPS equation for general pressure and for the axially symmetric ansatz (\ref{axi-sym}) (relevant, e.g., for neutron stars), which has a simple physical interpretation. Indeed, the equation reads
\be
\varepsilon_6 (r)= \varepsilon_0(r) + P
\ee
or, in words, {\it "the repulsive radial force per area equals the attractive radial force per area plus the pressure"}. For $P=0$, it just expresses the exact balance between repulsive and attractive forces at nuclear saturation, whereas in the limit of large pressure it shows that the repulsive force dominates, explaining the stiff character of the equation of state in that limit.

As a result of the above,
it is, therefore, reasonable to relate the two terms to the $\omega$ and $\sigma$ meson, respectively.
Schematically we may write
\be
\mathcal{L}_{BPS} =  - \pi^4 \lambda^2 \mathcal{B}_\mu \mathcal{B}^\mu- \nu^2 \mathcal{U} =
\mathcal{L}_\omega (U) + \mathcal{L}_\sigma(U).
\ee
Here, only  the pionic degrees of freedom (chiral $SU(2)$ fields) are present explicitly, whereas the (or at least some) effects of the  $\omega$ and $\sigma$ mesons are related to specific terms in the action. In other words, the $\omega$ and $\sigma$ mesons are hidden in the (nonlinear) Skyrme model action, and their effects are unravelled by studying the properties of particular solutions, similarly to the baryons themselves, which, too, are absent in the action and become visible only on the level of (solitonic) solutions as coherent superpositions of pion fields. In other words, both 
baryons (and atomic nuclei) and the $\omega$ and $\sigma$ mesons are realized in the model as emergent objects in a nonlinear pionic fluid.

Let us remark that the exact balance between attractive and repulsive forces in the BPS Skyrme model will be destroyed if we add the perturbative part to the action, that is, extend the model to the near-BPS Skyrme model (or include the semiclassical corrections). However, one can control this transition using the small $\epsilon$ parameter in the full action. 

As shown already, the sextic term leading to repulsion is responsible for the equivalence of the BPS Skyrme model and the Walecka model at high density (pressure), which allowed us to establish a precise relation between the Skyrme model parameters and parameters of the $\omega$ meson.  This quantitative relation is possible because the sextic term and the omega meson give the leading contribution at large density in the two models. 
For the potential term, though, a quantitative relation to the $\sigma$ meson is not so obvious, because their contribution is subleading at large densities, whereas at lower densities the $\sigma$ meson contributions in the Walecka model mix with other contributions (e.g., in-medium  fermion contributions). 

Finally, let us recall that the potential term explicitly breaks chiral symmetry in the Skyrme model EFT, therefore it should also be related to the (spontaneous or dynamical) chiral symmetry breaking in the underlying fundamental theory, i.e., QCD, and should, therefore, depend on the corresponding order parameter (the quark condensate $\langle \bar q q\rangle$; we remark that in the Walecka model there exists a linear relation between the nucleon condensate $\langle \bar\psi \psi \rangle$ and the $\sigma$ meson VEV). The BPS Skyrme model without potential, on the other hand, which should correspond to the case without chiral symmetry breaking, behaves completely differently at low densities (e.g., there is no nuclear saturation), and only in the limit of infinite density (infinite pressure) the BPS Skyrme model approaches the behavior of the pure sextic model without potential. This implies that for finite density (finite baryon chemical potential) chiral symmetry remains broken, and no phase transition to a chirally symmetric phase occurs. 

Let us also notice that other types of phase transitions are absent, too, in the BPS Skyrme model. There is, e.g., no phase transition of a topological type - skyrmions always remain skyrmions, and no fractional topological state, for example half-skyrmion state, is created when the pressure is increased \cite{rho-med}, \cite{half1}, \cite{half2}. The creation of such phases probably requires the inclusion of the perturbative (i.e., the non-BPS) part of the full near-BPS action. The details of the transition to these new topological phases should, then, be related to the mutual strength of BPS and non-BPS parts of the full effective theory. 
It is interesting to note, however, that the introduction of an external magnetic field may lead to topological phase transitions, at least in 2+1 dimensions \cite{magnetic}.
Also a phase transition to, e.g., quark matter cannot be described within the BPS Skyrme model alone. In other words, matter described by the BPS Skyrme model is always in a hadronic phase, and the only phase transition is the one between a gaseous hadronic phase below nuclear saturation and a liquid hadronic phase at and above nuclear saturation. It is interesting to note that this is precisely equivalent to the liquid-gas phase transition of nuclear matter, thus exactly reproducing the conjectured phase diagram of QCD at zero temperature for not too high values of the baryon chemical potential (close to nuclear saturation), see, e.g., \cite{fuku-hatsu}.            

%%%%%%%%%%%%%%%%%
\section{BPS Skyrme model as a perfect fluid}
%%%%%%%%%%%%%%%%%%%
A crucial property of the BPS Skyrme model was the fact that it has the energy momentum tensor of a perfect fluid, and that the static energy functional is invariant under SDiff transformations on physical space. Here we want to show briefly that the relation goes much further and that, at least formally, the action of the BPS Skyrme model is equivalent to the action of a field theoretic description of perfect fluids in an Eulerian formulation \cite{field-fluid}. There exist two main formulations of fluid mechanics, namely the Lagrangian formulation, where the dynamical variables are given by the particle trajectories (for finitely many particles) or by the fluid element trajectories (continuum limit), and the Eulerian formulation, where the dynamical variables have a more collective character and are provided by the (particle or mass) density $\rho$ and by the fluid velocity $\vec v$ (see, e.g., \cite{jackiw-fluid}). For finitely many particles, the d.o.f. in the Lagrangian formulation are the $N$ particle trajectories $\vec X_n (t)$, ($n=1,\ldots ,N$), and the corresponding mass density in  
the Eulerian formulation is $\rho (t,\vec x) = m\sum_n \delta^{(3)} (\vec X_n (t) - \vec x)$. In the continuum limit, the discrete particle label $n$ is replaced by three continuous labels $y^a$, $a=1,2,3$ required to label all fluid elements in three-dimensional space, and the corresponding dynamical variables are $\vec X(t,\vec y)$ in the Lagrangian formulation and
\bea
\rho (t,\vec x) &=&\rho_0 \int d^3 y \; \delta^{(3)} \left(\vec X (t,\vec y) - \vec x\right) , \nonumber \\
\vec v(t,\vec x) &=& \rho^{-1} \vec j \quad \mbox{where} \quad \vec j =  \rho_0 \int d^3 y \; \dot{\vec X} \delta^{(3)} \left( \vec X (t,\vec y) - \vec x\right)
\eea
in the Eulerian formulation. Here, the fluid element labels $y^a$ may be identified with the comoving coordinates of the fluid. For later convenience, we prefer to interpret $\rho$ as a particle number density (not a mass density). In addition, we prefer to include the dimensions of the $y^a$ into $\rho_0$, such that the $y^a$ are dimensionless, which makes $\rho_0$ dimensionless, too. We further remark that actions in the Lagrangian formulation, based on the $X^i (t, y^a)$ require integrations over $y^a$, i.e., not over physical space. Actions for the Eulerian formulation, on the other hand, include integrals over physical space $x^i$. The disadvantage of an action principle in the Eulerian formulation based on the dynamical fields $\rho$ and $\vec v$ is that the constraints required by hydrodynamics (particle number conservation, etc.) require the introduction of Lagrange multipliers, which complicates the analysis. Recently, however, a field theoretic version of the Eulerian formulation of fluid dynamics gained support \cite{field-fluid}, where the constraints are satisfied identically, evading thereby the necessity of Lagrange multipliers. In this field theoretic version, the comoving coordinates $y^a$ are promoted to the dynamical variables of the theory. Indeed, for a regularly flowing fluid the flow function $x^i = X^i (t,y^a)$ has an inverse $y^a = \phi^a (t,x^i)$ which allows to express the density like
\be
\rho (t,\vec x) = \rho_0  D \quad \mbox{where} \quad D\equiv \mbox{det} \left( \frac{\partial \phi^a}{\partial x^i}\right) .
\ee 
Here we assumed an Euclidean target space and cartesian coordinates with a volume form equal to one, but it is convenient to allow for non-cartesian coordinates (and, eventually, for a curved target space) with volume form $\Omega (\phi^a)d\phi^1 d\phi^2 d\phi^3$, leading to
\be
D = \Omega (\phi^a)\;   \mbox{det}  \left( \frac{\partial \phi^a}{\partial x^i}\right) .
\ee
The determinant $D$ has the further expression $D^2=\Omega^2 \; \mbox{det} (\partial_i\phi^a\partial^i \phi^b)$ which is useful, because it immediately allows for the relativistic generalization
\be
D^2 = \Omega^2\;  \mbox{det} \left( \frac{\partial \phi^a}{\partial x^\mu}\frac{\partial \phi^b}{\partial x_\mu} \right) .
\ee  
Finally, there exists a third expression for $D$, $6D^2 = {\cal N}^\mu {\cal N}_\mu$, in terms of the particle number current
\be
{\cal N}^\mu = \Omega \epsilon^{\mu\nu\rho\sigma} \epsilon_{abc} \partial_\nu \phi^a\partial_\rho \phi^b\partial_\sigma \phi^c .
\ee
It is now easy to find the relativistic generalizations of $\vec v$ and $\rho$. The velocity is replaced by the four-velocity $u^\mu$. The $\phi^a$ are the comoving coordinates which do not change along the flow, which implies $u^\mu \partial_\mu \phi^a =0$. $u^\mu$ must, therefore, be proportional to the particle number current ${\cal N}^\mu$, and the condition $u^\mu u_\mu =1$ leads to
\be 
u^\mu = \frac{{\cal N}^\mu}{\sqrt{{\cal N}^\nu {\cal N}_\nu }} = \frac{1}{\sqrt{6}D} \Omega \epsilon^{\mu\nu\rho\sigma} \epsilon_{abc} \partial_\nu \phi^a\partial_\rho \phi^b\partial_\sigma \phi^c ,
\ee
and the particle number density is defined by
\be \label{part-dens}
{\cal N}^\mu = \rho u^\mu \quad \Rightarrow \quad \rho = \sqrt{{\cal N}^\mu {\cal N}_\mu }  = \sqrt{6} D = \sqrt{6}\;  \Omega 
\left(  \mbox{det} \left( \frac{\partial \phi^a}{\partial x^\mu}\frac{\partial \phi^b}{\partial x_\mu} \right)  \right)^\frac{1}{2}.
\ee
Observe that particle number conservation $\partial_\mu {\cal N}^\mu=0$ (covariant conservation $\nabla_\mu {\cal N}^\mu =0$ in the general-relativistic case) is now an identity and does not require Lagrange multipliers.

In this setting, a perfect fluid action is defined by choosing a Lagrange density $F$ depending on $\phi^a ,\partial_\mu \phi^a$, etc. The simplest choice assumes that $F$ depends on the target space variables only via the scalar $\rho$, but more general perfect fluids may depend on further thermodynamic variables $h_A (\phi^a)$. Here we shall permit at most one further thermodynamical variable $h(\phi^a)$, i.e., $F=F(\rho ,h(\phi^a))$. The action just reads (we momentarily assume a general metric, for convenience)
\be
S =  \int d^4 x \sqrt{|g|} F(\rho ,h)
\ee
and leads to the energy-momentum tensor of a perfect fluid \cite{field-fluid},
\be
T^{\rho\sigma} \equiv -2|g|^{-\frac{1}{2}}\frac{\delta}{\delta g_{\rho\sigma}} S = (p+\varepsilon )u^\rho u^\sigma - pg^{\rho\sigma}
\ee
where
\be 
\varepsilon = -F(\rho ,h) \; ,\quad p = \rho \frac{\partial \varepsilon}{\partial \rho} - \varepsilon .
\ee
A particularly simple case occurs for $F=F(\rho)$. Then, both $\epsilon = \epsilon (\rho)$ and $p=p(\rho)$ are functions of the particle density $\rho$ only, and, upon eliminating $\rho$, an equation of state $\epsilon = \epsilon(p)$ may be found. Such fluids are called "barotropic". In the general case $F=F(\rho ,h)$, both $\epsilon$ and $p$ are functions of the two thermodynamic variables $\rho$ and $h$. A natural choice is $h=s(\phi^a)$, where $s$ is the entropy per particle. Observe that the entropy current ${\cal S}^\mu \equiv s{\cal N}^\mu$ is conserved identically ($\partial_\mu {\cal S}^\mu =0$ or $\nabla_\mu {\cal S}^\mu =0$, respectively).  

It is now quite obvious how to relate the perfect fluid field theory sketched above to the BPS Skyrme model. We just have to identify the Skyrme field $U$ (the fields $\xi, \Theta ,\Phi$, see (\ref{Sk-fields})) with the three scalar functions $\phi^a$ of the fluid. This identification is formal (only possible locally), because the $\phi^a$ take values in $\mathbb{R}^3$ (or a subspace thereof), whereas the Skyrme field takes values in SU(2) (or, equivalently, in $\mathbb{S}^3$). If we further assume that the volume form $\Omega$ is - at least locally - the volume form on $\mathbb{S}^3$, $\Omega = \sin^2 \xi \sin \Theta d\xi d\Theta d\Phi$, then the baryon density $\rho_B$ of the Skyrme model formally coincides with the particle number density $\rho$ of the perfect fluid, and the baryon current ${\cal B}^\mu$ coincides with the particle number current ${\cal N}^\mu$. Globally, the two currents are, of course, different. The topology of $\mathbb{S}^3$ guarantees, e.g., that the resulting baryon number $B$ is always an integer, which is not true for the particle number current. If we accept this formal analogy, then the lagrangian of the BPS Skyrme model is related to the fluid lagrangian
\be
F = -\lambda^2 \pi^4 \rho^2 -\nu^2 {\cal U}(\phi^a)
\ee
where $\rho$ is the particle density and the potential ${\cal U}$ corresponds to a further "thermodynamical variable". While the identification of $\rho$ with the baryon density $\rho_B$ is obvious, it is not so clear what thermodynamical variable should correspond to the potential. The identification of ${\cal U}$ with the entropy per particle is not plausible, because the nuclear matter which the BPS Skyrme model is supposed to describe is essentially at zero temperature. Finally, the two cases of the BPS Skyrme model with the step function potential and without potential lead to lagrangians $F(\rho)$ which only depend on $\rho$, and, therefore, correspond to barotropic fluids. BPS Skyrme models with genuine, $U$-dependent potentials, on the other hand, correspond to non-barotropic (i.e., baroclinic) fluids.

%%%%%%%%%%%%%%%%%%%%%%%%%%%%%%%%%%%%%%%%%
\section{Conclusions}
%%%%%%%%%%%%%%%%%%%%%%%%%%%%%%%%%%%%%%%%%
The main result of the present paper is the introduction and analytical description of the baryon chemical potential for the BPS Skyrme model, which is one necessary ingredient for a full understanding of the thermodynamical properties of skyrmions as nuclear matter at zero temperature. We found the especially simple result that the baryon chemical potential is just the baryon charge density multiplied by a constant. As for other global quantities in the BPS Skyrme model (the total energy and volume), also the mean-field baryon chemical potential can be analytically obtained (as a function of the pressure $P$) by a certain integral (average) over the target space.  In other words, the BPS Skyrme model realises to the very extreme the concept of a geometric model of matter. Indeed, all global quantities (energy, volume, mean-field energy density, mean-field chemical potential) can be computed without the knowledge of particular solutions. We also confirmed that the baryon chemical potential obeys all required thermodynamical relations.  

The complete analysis of the thermodynamics of the BPS Skyrme model at $T=0$ also allowed to conclude that the model is equivalent to the Walecka model (as well as a vector interaction enhanced bag model)) in the high density regime. At densities close to the saturation point, however, the behavior of the BPS Skyrme model is more involved. Depending on the form of the potential or, more precisely, on how it approaches the vacuum, one may have a kind of bag model (with hadronic degrees of freedom) behavior with a non-zero energy density at zero pressure or, instead, a zero energy density onset. The first case occurs for  potentials whose near-vacuum dependence is $\mathcal{U} \sim \xi^\alpha$, $\alpha <6$. This case is still qualitatively similar to the Walecka model in that both models show nuclear saturation at the saturation density. The main difference is that in the Walecka model there is a region of negative pressure (long-range attractive forces) below but close to nuclear saturation density, whereas pressure is exactly zero below nuclear saturation in the BPS Skyrme model, based on its classical soliton solutions.
This behavior should, however, change once further terms of the near-BPS Skyrme model are included. It is, for instance, known that the standard non-linear sigma model term ${\cal L}_2 = \lambda_2 L_\sigma L^\sigma $ induces attractive long-range forces between skyrmions in some (attractive) spin-isospin channels \cite{man-sut-book}. Also the inclusion of binding energies due to quantum corrections (semi-classical quantization) of the skyrmion energies should change this behavior in a similar fashion. 

It is important to notice that the equivalence with the Walecka and related models is obtained in the mean-field limit. The BPS Skyrme model, on the other hand, also provides in a natural way a description beyond mean-field theory, that is, on the level of local, field theoretical quantities (energy density, baryon number density). In the asymptotic large density regime, where the exact equivalence is established, the local field theoretical computations agree with their mean-field approximations.
In a lower density regime, however, where $\mu$ and $\bar{\mu}$ are different, we think that the full field-theoretic non-mean-field quantities (baryon chemical potential, energy density, etc.) should be used for a more precise description. 

Furthermore, we found some evidence that the BPS Skyrme model can be interpreted as an $\omega$-meson dominated model of nucleons,  where the $\omega$-mesons are hidden in the form of the action (or, more precisely, in the derivative term used in the action - baryon current squared) rather than included as an effective low energy field. We remind the reader that, due to the absence of the quadratic term ${\cal L}_2 = \lambda_2 L^\mu L_\mu$, there are no propagating pions and, therefore, no forces mediated by pions in the BPS Skyrme model. This simply means that, as stated repeatedly, the BPS Skyrme model by itself {\it cannot be considered a complete low-energy effective theory} for nuclear physics or strong interactions. A good candidate for such a low-energy effective theory is, in our opinion, the near-BPS Skyrme model
\be
{\cal L} = {\cal L}_{\rm BPS} + \epsilon \left(  {\cal L}_2 +  {\cal L}_4 + \tilde {\cal L}_0 \right)
\ee
(here ${\cal L}_4 = \lambda_4 [L_\mu , L_\nu ]^2$ is the Skyrme term, and $\tilde {\cal L}_0$ is a further potential), where $\epsilon$ is assumed to be small in the sense that ${\cal L}_{\rm BPS}$ provides the main contributions to soliton masses and is dominant in regions of sufficienty large baryon density, such that the unique properties of the BPS submodel are essentially preserved in this regime. Close to the vacuum, on the other hand, ${\cal L}_2$ always dominates over ${\cal L}_6$, and long-range forces mediated by pions are, therefore, present in the near-BPS extension. It is of some interest to note that the BPS Skyrme model does reproduce the forces related to the $\omega$ and $\sigma$ mesons - whose fields do not appear in the action - while it does not include the pionic forces, although its action is expressed entirely in terms of pion fields.

As already mentioned, there are not many results concerning the baryon chemical potential in the Skyrme theory. One comparison, however, may be made with results found by means of the AdS/CFT correspondence and the Sakai-Sugimoto model \cite{ss}, which is a holographic (large $N_c$ relevant) version of a Skyrme type model. It has been found that for the two flavour case asymptotically the baryon density behaves as for a free fermion gas $\rho_B \propto \mu^3$ \cite{ram}, which strongly differs form our result. In spite of that, there is a qualitative similarity between this holographic computation and our approach. In both models, the baryon density is always a function of the space coordinates at any finite value of the chemical potential. Such an inhomogeneous configuration tends to a homogeneous one only in the infinite density (chemical potential) limit \cite{ram}. On the other hand, it is possible to get an asymptotically linear relation between the chemical potential and baryon density in a holographic set-up. For this, one needs the usual Maxwell action instead of the Born-Infeld one \cite{nog}. Adding more scalar fields may, however, change this linear relation. 

There are several obvious directions in which the current investigations should be continued. First of all, the knowledge of the baryon chemical potential for the BPS Skyrme model is essential if one wants to apply the model for a complete description of neutron stars. In principle, the core of neutron stars, described by the BPS Skyrme model, may be surrounded by a skin with more usual matter with a known equation of state. The obvious condition for a transition from the dense hadronic phase (BPS Skyrme action) to the skin phase is the equivalence of the chemical potentials. This should lead to a modification of the mass-radius relation for the low massive stars with perhaps an appearance of a minimal neutron star mass \cite{lukasz}. This, however, will be modified already by the inclusion of the non-BPS part of the full near BPS Skyrme action. Indeed, as one approaches the outer region of a neutron star in the BPS Skyrme model, the matter Skyrme field tends to its vacuum value. But close to the vacuum, the perturbative terms in the chiral Lagrangian dominate over the BPS part. Hence, in this regime they cannot be omitted. This points towards another important issue which should be understood, namely the generalization of the thermodynamical description of the BPS Skyrme theory to its near-BPS extension, as well as its application to nuclear phenomenology. Let us notice that, except for the step-function potential case, one should use the baryon chemical potential rather than its mean-field approximation. In fact, for non-flat potentials it is known that local quantities, especially for heavy neutron stars, change a lot if computed in the mean-field limit (compare pressures and energy densities inside neutron stars in \cite{star2}).

Another important quantity which has to be understood is the isospin chemical potential (there already exist some proposals on how to treat this issue in the Skyrme framework \cite{iso}). This would allow for a complete description of skyrmionic nuclear matter at zero temperature.  
%%%%%%%%%%%%%%%%%%%%%%%%%%%%%%%%%%%%%%%%%
\section*{Acknowledgement}
%%%%%%%%%%%%%%%%%%%%%%%%%%%%%%%%%%%%%%%%%
The authors acknowledge financial support from the Ministry of Education, Culture, and Sports, Spain (Grant No. FPA2011-22776), the Xunta de Galicia (Grant No. INCITE09.296.035PR and Conselleria de Educacion), the Spanish Consolider-Ingenio 2010 Programme CPAN (CSD2007-00042), and FEDER. This work was supported by the Polish National Science Center (NCN) under grant number UMO-2013/09/B/ST2/01560 (T.K.). AW was supported by the Polish KNOW FOCUS Jagiellonian University grant (2015).  
CN thanks the Spanish Ministery of
Education, Culture and Sports for financial support (grant FPU AP2010-5772).
AW thanks L. Bratek and J. Jachola for discussion.

\end{document}